\documentclass[aps,prd,twocolumn,showpacs,showkeys,superscriptaddress]{revtex4-1}
\usepackage{latexsym}
\usepackage{amssymb}
\usepackage{amsmath}
\usepackage{amscd}
\usepackage{amsthm}
\usepackage{graphicx}
\usepackage{textcomp}
\usepackage{colortbl}
\usepackage{hyperref}
\usepackage[font={footnotesize,it}]{caption}
\usepackage{multirow}
\usepackage[T1]{fontenc}
\usepackage{ae,aecompl}

\begin{document}

\title{An exact solution of the observable universe in Bianchi V space-time}
\author{Rajendra Prasad}
\email[]{drrpnishad@gmail.com}
\affiliation{Department of Physics, Galgotias College of Engineering and Technology, Greater Noida 201310, India}
\author{Manvinder Singh}
\email[]{manvindernps@gmail.com}
\affiliation{Department of Applied Science, G. L. Bajaj Institute of Technology and Management, Greater Noida 201306, India}
\author{Anil Kumar Yadav}
\email[]{abanilyadav@yahoo.co.in}
\affiliation{Department of Physics, United College of Engineering and Research,Greater Noida - 201310, India}
\author{A. Beesham}
\email[]{abeesham@yahoo.com}
\affiliation{Department of Mathematical Sciences, University of Zululand, Kwa-Dlangezwa 3886, South Africa\\
Faculty of Natural Sciences, Mangosuthu University of Technology, Umlazi 4026, South Africa}

\begin{abstract}
In this paper we investigate an observable universe in Bianchi type V space-time by taking into account the cosmological constant as the source of energy. We have performed a $\chi^{2}$ test to obtain the best fit values  of the model parameters of the universe in the derived model. We have used two types of data sets, viz: i) 31 values of the Hubble parameter and ii) the 1048 Phanteon data set of various supernovae distance moduli and apparent magnitudes. From both the data sets, we have estimated the  current values of the Hubble constant, density parameters  $(\Omega_{m})_{0}$  and $(\Omega_{\Lambda})_{0}$. The present value of deceleration parameter of the universe in derived model is obtained as $q_{0} = 0.59^{+0.04}_{-0.03}$ and $0.59^{+0.02}_{-0.03}$ in accordance with $H(z)$ and Pantheon data respectively. Also we observe that there is a signature flipping in the sign of deceleration parameter from positive to negative and transition red-shift exists. Thus, the universe in derived model represents a transitioning universe which is in accelerated phase of expansion at present epoch. We have estimated the current age of the universe $(t_{0})$ and present value of jerk parameter $(j_{0})$. Our obtained values of $t_{0}$ and $j_{0}$ are in good agreement with its values estimated by Plank collaborations and WMAP observations.
\end{abstract}



\maketitle
\section{Introduction}\label{1}
In the first decade of twenty one century, there has been developments in cosmology that have reinterpreted the cosmological constant $(\Lambda)$. Firstly, the idea of inflation gave cosmology a whole new view upon the first split second of our universe. A key ingradient in the inflationary model is the behaviour of model that have a cosmological constant like behaviour. Secondly, recent astrophyical observations indicate that we live in an accelerated universe. The inclusion of $\Lambda$ in Einstein's field equation can give rise to such behaviour as we will show in in this paper. Also, the cosmological observations leaded by SN Ia groups had confirmed that the universe is in accelerated expansion phase at present epoch \cite{ref1,ref2,Perlmutter/1999}. This acceleration in the universe may driven by an exotic type of unknown fluid that have positive energy density and huge negative pressure. This fluid is usually known as Dark Enegry (DE) but its nature is still unknown. The most suitable candidate of this DE is the $\Lambda$. However, there is a huge dissimilarity in the value of $\Lambda$ predicted by observations and particle physics ground that leads tuning problem. In Refs. \cite{Sahni/2003a,Padmanabhan/2003}, the authors have addressed fine tuning problem of $\Lambda$ and given a clue to solve tuning problem associated with $\Lambda$ by assuming its dynamical character with respect to time or red-shift. The DE is less effective in early universe but it dominates the present universe and since it does not interact with the baryonic matter, hence it makes difficult to detect. In Ma et al. \cite{Ma/1999}, the authors have studied non interaction of DE and baryonic matter with the cosmic expansion history and the growth rate of cosmic large scale structure. Several theoretical models based on the phenomenon of late time acceleration in the universe have been proposed in last two decades mostly; post supernovae observations \cite{ref4}-\cite{ref14}. In recent past, the hybrid scale factor which can mimic the cosmic transitive behaviour of the universe from early decelerated phase to late time accelerated phase has been investigated \cite{Mishra15,Mishra17,Mishra18a}. The cosmological models in a non-interacting two fluid scenario such as the usual DE and electromagnetic field have been studied \cite{Ray19,Mishra19a}. The outcome of this research is that the DE dominates the universe at present epoch and derives the late time acceleration of the universe. Some important applications of DE in anisotropic space time are given in Refs. \cite{Kumar/2011mpla,Amirhashchi/2011plb}.

In recent times, Bianchi type V cosmological models have commended the attention of several cosmologists due to the facts that its dynamical behaviour can be for more general in comparison to FRW and Bianchi type I cosmological models. The Bianchi V space-time is still less complicated than Bianchi type II, VI and IX. Further it is observed that Bianchi type V space time permits small anisotropic at fixed stages in the evolution of the universe \cite{Singh/2009bjp}. So, Bianchi type V cosmological models create more interest and due to its some specific properties, these models turn into a suitable model for study of the universe. One of the important outcome of CMB observations is that it favors the existence of anisotropic phase which approaches isotropic one at later times. Therefore, it make sence to study the universe with anisotropic background in presence of cosmological terms $\Lambda$. Note that in recent years, Bianchi type cosmological models are playing invaluable role in observational cosmology as the WMAP observational data \cite{Hinshaw/2003,Hinshaw/2007,Hinshaw/2009} has confirmed an addition to standard $\Lambda$CDM model that bears a likeness to the Bianchi morphology \cite{Jaffe/2005,Jaffe/2006,Jaffe/2006a,Campanelli/2006}. Therefore, in spite of inflation, the universe has slightly anisotropic special geometry that leads to a non-trivial isotropization history of universe due to the presence of an anisotropic energy source. Some sensible research in Bianchi type V space-time are given in Refs. \cite{Collins/1974,Coles/1994,Maartens/1978}. In particular, Collins \cite{Collins/1974} has investigated simplest perfect fluid cosmological models in Bianchi V space-time which possess a singularity. One of the interesting result of this study is that this model consist of two disjoint regions in  which the matter density is non-zero, separated by a Cauchy horizon on which the matter density is null. In Coles and Ellis \cite{Coles/1994}, authors have discussed the case of an open universe and investigated that observations support a model of the universe which has low density. Maartens and Nel \cite{Maartens/1978} had investigated some new and exact solution in spatially homogeneous Bianchi V space-time that admit a non-vanishing magnetic field. In our previous work \cite{Goswami/2020mpla}, we have studied the kinematics and fate of Bianchi V universe by taking into account the dynamical nature of DE while in this paper, we have considered cosmological constant as source of energy. THe mechanism of obtaining Hubble's function $H(z)$ are altogether different from $H(z)$ given in Ref. \cite{Goswami/2020mpla}. 

In this work, we have investigated a Bianchi type-V model of the universe in which baryons also have pressure. It has been stated that ``All of observational cosmology is the search for two numbers: i) Hubble parameter (HP) and ii) deceleration parameter (DP) $H_0$ and $q_0$" \cite{ref33}. In the present scenario, higher derivatives of the scale factor, such as the jerk parameter $(j_0)$, snap parameter $(s_0)$ and lerk parameter $(l_0)$ do play a role. A successful cosmological model will be one in which these parameters fit best with the observational inputs. Keeping this as our motto, we have performed a $\chi^{2}$ test to obtain the best fit values of the model parameters of the universe in our derived model that leads to good consistency of the theoretical model with observations. We have used two types of data sets: i) A data set of 31 Hubble parameter values and  ii) The 1048 Phanteon data set of various supernovae distance moduli and apparent magnitudes. From both data sets, we have estimated current values of the Hubble constant, density parameters  $(\Omega_{m})_{0}$ and $(\Omega_{\Lambda})_{0}$. The present value of deceleration parameter of the universe in derived model is obtained as $q_{0} = 0.59^{+0.04}_{-0.03}$ and $0.59^{+0.02}_{-0.03}$ in accordance with $H(z)$ and Pantheon data respectively. This values of $q_{0}$ is very close to its empirical value obtained by Cunha et al \cite{Cunha/2009}. Also we observe that there is a signature flipping in the sign of deceleration parameter from positive to negative and transition red-shift exists. Thus, the universe in derived model represents a transitioning universe which is in accelerated phase of expansion at present epoch. We have estimated the current age of the universe $(t_{0})$ and present value of jerk parameter $(j_{0})$. Our obtained values of $t_{0}$ and $j_{0}$ are in good agreement with its values estimated by Plank collaborations and WMAP observations. 

The paper is organized as follows: Section \ref{1} deals with the straightforward description of the various investigations and their  results in the thrust area of the paper. In section \ref{2}, the model and its basic equations are presented. Section \ref{3} is devoted to data and likelihoods. Some physical parameters and properties of the universe in our derived model are described in section \ref{4}. Finally, we have given conclusion of this research in Section \ref{5}.    
\section{The model and Basic equations}\label{2}
The anisotropic and homogeneous Bianchi type V space-time is read as
\begin{equation}
\label{metric}
ds^{2} = dt^{2}-X^{2}(t)dx^{2}- exp(2\alpha x)\left[Y^{2}(t)dy^{2}- Z^{2}(t)dz^{2}\right]
\end{equation}
where $X(t)$, $Y(t)$ and $Z(t)$ are scale factors along the $x$, $y$ and $z$ axes. The exponent $\alpha$ is an arbitrary constant.\\
Einstein's field equation is read as
\begin{equation}
\label{fe}
R_{ij}-\frac{1}{2}Rg_{ij}-\Lambda g_{ij} = 8\pi G T_{ij}
\end{equation}
where $R$ is the Ricci scalar and $T_{ij}$ is energy-momentum tensor of perfect fluid.\\ 
The energy-momentum tensor $(T_{ij})$ of the perfect fluid is given as
\begin{equation}
\label{fe1}
T_{ij} = (\rho + p)v_{i}v_{j} - pg_{ij}
\end{equation}
where $v^{i}$ is the four velocity vector which satisfies $v^{i}v_{i} = 1$. $\rho$ and $p$ are the energy density and pressure of the perfect fluid.\\ 
Solving (\ref{fe}) with the space-time metric (\ref{metric}), we get the following system of equations
\begin{eqnarray}
\frac{\ddot{Y}}{Y}+\frac{\ddot{Z}}{Z}+\frac{\dot{Y}\dot{Z}}{YZ} -\frac{\alpha^{2}}{X^{2}}&=& -8\pi G p +\Lambda  \label{fe2} \\
\frac{\ddot{X}}{X}+\frac{\ddot{Z}}{Z}+\frac{\dot{X}\dot{Z}}{XZ}-\frac{\alpha^{2}}{X^{2}}&=& -8\pi G p +\Lambda \label{fe3} \\
\frac{\ddot{X}}{X}+\frac{\ddot{Y}}{Y}+\frac{\dot{X}\dot{Y}}{XY}-\frac{\alpha^{2}}{X^{2}}&=& -8\pi G p +\Lambda \label{fe4} \\
\frac{\dot{X}\dot{Y}}{XY}+\frac{\dot{Y}\dot{Z}}{YZ}+\frac{\dot{Z}\dot{X}}{ZX}-\frac{3\alpha^{2}}{X^{2}}&=& 8\pi G \rho +\Lambda \label{fe5}\\
2\frac{\dot{X}}{X}-\frac{\dot{Y}}{Y}-\frac{\dot{Z}}{Z} = 0 & \Rightarrow & X^{2} = \kappa_{1}YZ \label{fe5-1}
\end{eqnarray}
where $\kappa_{1}$ is constant of integration and we have taken $\kappa_{1} = 1$ without loss of generality.\\
Equations (\ref{fe2})-(\ref{fe4}) lead to the following system of equations 
\begin{equation}
\label{fe6}
\frac{\ddot{X}}{X}-\frac{\ddot{Y}}{Y}+\frac{\dot{X}\dot{Z}}{XZ}-\frac{\dot{Y}\dot{Z}}{YZ} = 0
\end{equation}
\begin{equation}
\label{fe7}
\frac{\ddot{Y}}{Y}-\frac{\ddot{Z}}{Z}+\frac{\dot{X}\dot{Y}}{XY}-\frac{\dot{X}\dot{Z}}{XZ} = 0
\end{equation}
\begin{equation}
\label{fe8}
\frac{\ddot{Z}}{Z}-\frac{\ddot{X}}{X}+\frac{\dot{Y}\dot{Z}}{YZ}-\frac{\dot{X}\dot{Y}}{XY} = 0
\end{equation}
If $\xi$ is an arbitrary function of $t$, then equation (\ref{fe5-1})satisfies the following 
\begin{equation}
\label{s-1}
B = A\xi \;\;\; \& \;\; C = \frac{A}{\xi}
\end{equation}
where $\xi = \xi(t)$ relates to the anisotropy in the universe. \\

Using equation (\ref{s-1}) in any one of the equations (\ref{fe6}) - (\ref{fe8}), we obtain 
\begin{equation}
\label{s-2}
\frac{\ddot{\xi}}{\xi}-\frac{\dot{\xi}^{2}}{\xi^{2}}+\frac{\dot{3\xi}}{\xi}\frac{\dot{X}}{X} = 0
\end{equation}
After integration of equation (\ref{s-2}), we obtain
\begin{equation}
\label{s-3}
\frac{\dot{\xi}}{\xi} = \frac{K}{X^{3}}
\end{equation}
Now, the average scale factor $a(t)$ is defined as
\begin{equation}
\label{as}
a = (XYZ)^{\frac{1}{3}}
\end{equation}
Finally, the field equations of Bianchi type V space-time in term of average scale factor are read as 
\begin{equation}\label{fe9}
2\frac{\ddot{a}}{a}+ \frac{\dot{a}^{2}}{a^{2}}-\frac{\alpha^{2}}{a^{2}} = -8\pi G p +\Lambda -\frac{K^{2}}{ a^{6}}
\end{equation}
\begin{equation}\label{fe10}
3\frac{\dot{a}^{2}}{a^{2}}-3\frac{\alpha^{2}}{a^{2}}= 8\pi
G \rho +\Lambda + \frac{K^{2}}{ a^{6}}
\end{equation}
From equations (\ref{fe9}) and (\ref{fe10}), we observe that it is a system of two equations with three variables. Hence, one con not solve it in general. Therefor, to get an explicit solution, we have to adopt an additional but physically reasonable condition that is why we take well known barotropic equation of state which quantifies the relation between energy density $(\rho)$ and pressure $(p)$. $i. e.$
\begin{equation}
\label{es}
p = \omega \rho
\end{equation}     
where $ 0 \leq \omega \leq 1$ is known as equation of state parameter of perfect fluid.\\

Solving equations (\ref{fe9}), (\ref{fe10}) and (\ref{es}), we obtain the following differential equation
\begin{equation}\label{hub}
\frac{2\ddot{a}}{a}+\frac{(3 \omega +1) \dot{a}^2}{a^2}-\frac{K^2(\omega -1)}{a^6}-(3\omega+1)\frac{\alpha^{2}}{a^{2}} = \Lambda(\omega+1)
\end{equation}
Using the standard definition, $H = \frac{\dot{a}}{a}$ and $a = \frac{1}{1+z}$, equation (\ref{hub}) leads to 
\begin{widetext}   
\begin{equation}\label{hub1}
2(z+1) H H'-3 (\omega+1) H^2+\Lambda (\omega+1)+K^2 (\omega-1) (z+1)^6+(3\omega+1)\alpha^{2}(1+z)^{2}=0
\end{equation}
\end{widetext}
where $H'$ is the first order derivative of $H$ with respect to $z$.\\
Solving equation (\ref{hub1}), we obtain
\begin{widetext}
\begin{equation}\label{hub2}
 H = H_0  \sqrt{(\Omega_{m})_0 (1+z)^{3(\omega+1)}+(\Omega_{\alpha})_{0}(1+z)^{2}+(\Omega_{\sigma})_0(1+z)^6 +(\Omega_{\Lambda})_0}
\end{equation}
\end{widetext}
\begin{widetext}
where $$(\Omega_{m})_0= 1 -\frac{ \Lambda }{3 H_0^2}-\frac{K^2}{3 H_0^2} -\frac{\alpha^{2}}{3H_{0}^{2}},
  ~~ (\Omega_{\Lambda})_0=\frac{\Lambda}{3H_{0}^2}, ~~ (\Omega_{\alpha})_{0} = \frac{\alpha^{2}}{3H_{0}^{2}}
 ~~\&~~(\Omega_{\sigma})_0=\frac{K^2 }{3 H_0^2} $$
\end{widetext} 
Thus, we have
$$(\Omega_{m})_0 +(\Omega_{\Lambda})_0+(\Omega_{\sigma})_0+ (\Omega_{\alpha})_{0} = 1$$
where $H_{0}$ is the present value of Hubble's parameter.\\   
Now, the expressions for luminosity distance $(D_L)$ and distance modulus $(\mu)$ of any distant luminous object are determined as     
\begin{equation}
\label{eq9}
D_{L}(z)=\frac{(1+z)}{H_{0}}\int^z_0\frac{dz} {h(z)} ; \;\; h(z) = \frac{H}{H_{0}}
\end{equation}
and 
\begin{equation}\label{eq10}
\mu(z) = m_{b} - M = 5log_{10}\left(\frac{D_{L}(z)}{Mpc}\right) + \mu_{0}
\end{equation}
where $m_{b}$ and $M$ are apparent magnitude and absolute magnitude of any distant luminous object respectively. $\mu_{0}$ is the zero point offset.
\section{Data and likelihood}\label{3}
In this section, we use the observational $H(z)$ data and recent SN Ia Pantheon data. Also, we describe here the statistical methodology for constraining different model parameters of the derived universe.\\
\begin{itemize}
\item {\bf Observational Hubble Data (OHD)}: For Hubble H(z) data, we adopt $31~H(z)$ observational datapoints in the range of $0\leq z\leq 1.96$ taken table 1 of Ref. \cite{HA/2020PDU}. The cosmic chronometric (CC) technique is adopted to determined these uncorrelated data. There reason behind to take this data is the fact that OHD data obtained from
CC technique is model-independent. In fact, the most evolving galaxies based on the ``galaxy differential age'' method is used to determine this CC data \cite{Moresco/2016}.\\

\item {\bf Pantheon Data}: For the SN Ia data, we consider the recent Pantheon sample compiled in Scolnic et al. \cite{Scolnic/2018}. This Pantheon SN Ia catalogue is publicly available in \href{http://dx.doi.org/10.17909/T95Q4X}{Scolnic et al. 2018}.\\
\end{itemize}
\begin{figure}[ht!]
\begin{center}
\includegraphics[width=4.0cm,height=4.0cm,angle=0]{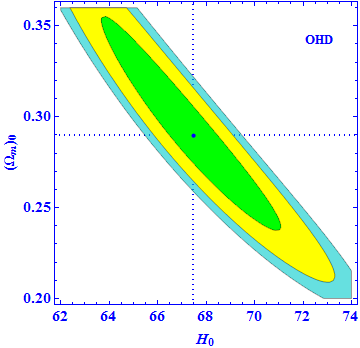}
\includegraphics[width=4.0cm,height=4.0cm,angle=0]{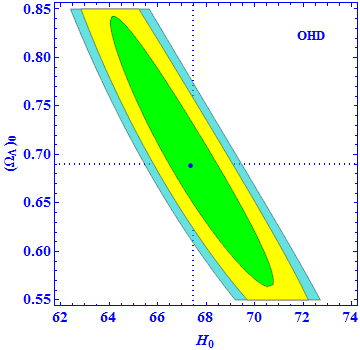} 
\caption{Two dimensional (2D) contours at $1\sigma$, $2\sigma $ and $3\sigma$ confidence regions by bounding the derived model with latest 31 observational Hubble data compiled from CC technique. The estimated values of $H_{0} = 67.46 \pm 1.2\;km\;s^{-1}\;Mpc^{-1}$, $(\Omega_{\Lambda})_{0} = 0.69 \pm 0.01$ and $(\Omega_{m})_{0} = 0.29 \pm 0.005$.}
\end{center}
\end{figure}
\begin{figure}[ht!]
\centering
\includegraphics[width=4cm,height=4cm,angle=0]{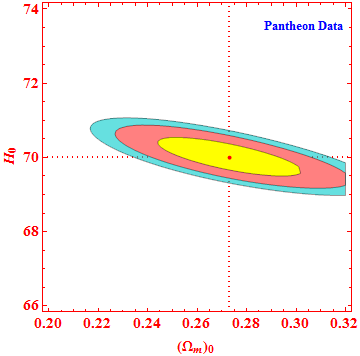} 
\includegraphics[width=4cm,height=4cm,angle=0]{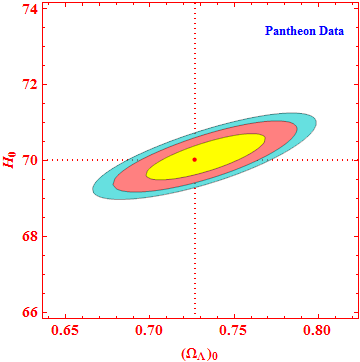}
\caption{Two dimensional (2D) contours at $1\sigma$, $2\sigma $ and $3\sigma$ confidence regions by bounding the derived model with Pantheon data. The estimated values of $H_{0} = 70.02 \pm 1.8\;km\;s^{-1}\;Mpc^{-1}$, $(\Omega_{\Lambda})_{0} = 0.727 \pm 0.014$ and $(\Omega_{m})_{0} = 0.273 \pm 0.03$.}
\end{figure}
For probing the model parameters, we have defined $\chi^{2}$ for the parameters with the likelihood given by $\varphi \propto e^{-\frac{\chi^2}{2}}$.
\begin{equation}
\label{chi-1}
\chi^{2}_{ OHD}=\sum_{i=1}^{31}\left[\frac{H(z_{i},\Psi)-H_{obs}(z_{i})}{\sigma_{i}}\right]^{2}
\end{equation}
where $\Psi$ and $\sigma_{i}$ represent the parameter vector and standard error in experimental values of the Hubble function $H$ respectively. \\

Now, we assume following uniform priors enforced on the model parameters for statistical analysis.
$$H_{0} \sim (50,80),\;\; (\Omega_{m})_{0} \sim (0,0.5),\;\&\; (\Omega_{\Lambda})_{0} \sim (0,1)$$
Figure 1 shows 2D contours at $68\%$, $95\% $ and 
$99\%$ confidence regions by bounding our model with latest 31 observational Hubble data. The estimated values of $H_{0} = 67.46 \pm 1.2\;km\;s^{-1}\;Mpc^{-1}$, $(\Omega_{\Lambda})_{0} = 0.69 \pm 0.01$ and $(\Omega_{m})_{0} = 0.29 \pm 0.005$.
Similarly, for the Pantheon data, we have
\begin{equation}
\label{chi-2}
\chi^{2}_{Pantheon}=\sum_{i=1}^{1048}\left[\frac{\mu(z_{i},\Psi)-\mu_{obs}(z_{i})}{\sigma_{i}}\right]^{2}
\end{equation}
where $\Psi$ and $\sigma_{i}$ are parameter vector and standard error in experimental values of $\mu(z)$ respectively.\\  
Figure 2 depicts 2D contours at $68\%$, $95\%$ and 
$99\%$ confidence regions by bounding the model under consideration with recent Pantheon data. The estimated values of $H_{0} = 70.02 \pm 1.8\;km\;s^{-1}\;Mpc^{-1}$, $(\Omega_{\Lambda})_{0} = 0.727 \pm 0.014$ and $(\Omega_{m})_{0} = 0.273 \pm 0.02$.\\
\section{Physical properties of the Model}\label{4}
\subsection{Deceleration parameter}
The deceleration parameter (DP) is read as
\begin{equation}
\label{q-0}
q = -\frac{a\ddot{a}}{\dot{a}^{2}} = -1+\frac{(1+z)H^{\prime}}{H}
\end{equation}
where $\dot{z} = -(1+z)H$.\\
The the expression of $q$ of our derived model of the universe is obtained as
\begin{widetext}
\begin{equation}
\label{q-1}
q = -1+\frac{3(1+\omega)(\Omega_{m})_{0}(1+z)^{3(1+\omega)}+6(\Omega_{\sigma})_{0}(1+z)^{6}+2(\Omega_{\alpha})_{0}(1+z)^{2}}{2[(\Omega_{m})_{0}(1+z)^{3(1+\omega)}+(\Omega_{\Lambda})_{0}+(\Omega_{\sigma})_{0}(1+z)^{6} + (\Omega_{\alpha})_{0}(1+z)^{2}]}
\end{equation}
\end{widetext} 
Therefore, the present value of DP $(q_{0})$ is computed by putting $z = 0$ in equation (\ref{q-1}) 
\begin{equation}
\label{q-2}
q_{0} = -1+\frac{3(1+\omega)(\Omega_{m})_{0}+6(\Omega_{\sigma})_{0}+2(\Omega_{\alpha})_{0}}{{2[(\Omega_{m})_{0}}+(\Omega_{\Lambda})_{0}+(\Omega_{\sigma})_{0}+(\Omega_{\alpha})_{0}]}
\end{equation}
Thus, the present values of DP of the universe in derived model are estimated as $q_{0} = -0.59^{+0.04}_{-0.03}$ and $q_{0} = -0.59^{+0.04}_{-0.03}$ fit with OHD and Pantheon data respectively. In Fig. 3, the best fit curve of $q$ at 
$68\%$ confidence level is shown. Note that the estimated value of $q_{0}$ in this paper is nicely tally with its value obtained by observational researches \cite{Cunha/2009,Prasad/2020,Sharma/2020ijgmmp}. Therefore, the universe in derived model is in good agreement with recent observations. Further, we observe that, the early universe was in decelerated phase of expansion while the current universe repels its ingredient matter/energy with acceleration. Hence, the universe in derived model represents a model of transitioning universe that have signature flipping at $z_{t} = 0.74\pm0.06$ and $z_{t} = 0.72\pm0.04$ with respect to OHD and pantheon data respectively. Some sensible researches for transition red-shift $(z_{t})$ are as follows: Farooq et al \cite{Farooq/2017} have estimated $z_{t}= 0.72\pm 0.05$ by using the 38 H(z) data points (Also see \cite{Farooq/2013} for the results obtained from the 28 H(z) data). Recently, in \cite{Amirhashchi/2019prd}, the authors have given a comparison between transition red-shift of the $\Lambda$CDM and $\Lambda$BI models. So, the obtained values of transition red-shift $(z_{t})$, in this paper, is very close to the previous results.\\
\begin{figure}[ht!]
\begin{center}
\includegraphics[width=4cm,height=4cm,angle=0]{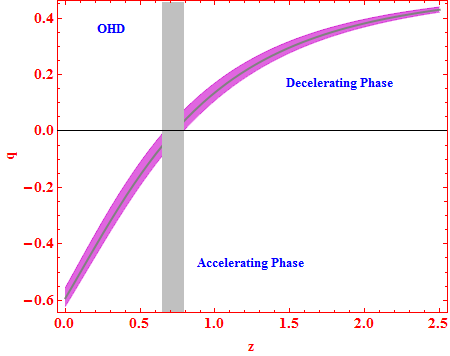} 
\includegraphics[width=4cm,height=4cm,angle=0]{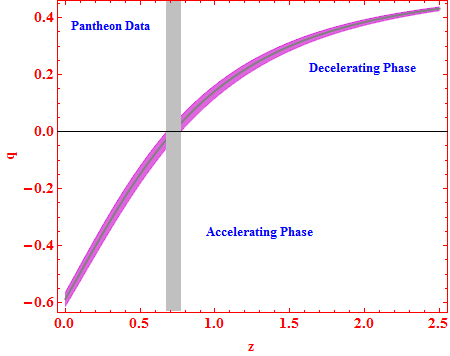}
\caption{Plots of $q(z)$ versus $z$ for the model parameters obtained from bounding the derived model with OHD (left panel) and Pantheon data (right panel). The transition red-shift is obtained as $z_{t} = 0.74\pm0.06$ and $z_{t} = 0.74\pm0.04$ for OHD and Pantheon data respectively.}
\end{center}
\end{figure}
\subsection{Age of the universe}
The age of the universe is obtained as
\begin{align}
\label{age-1}
& dt = -\frac{dz}{(1+z)H(z)}\Rightarrow 
\int_{t}^{t_{0}} dt  =  \int_{0}^{z}\frac{1}{(1+z)H(z)}dz
\end{align}
where $t_{0}$ denotes present age of the universe in our derived model.\\
Thus, the present age of the universe is read as
\begin{widetext}
\begin{equation}
\label{age-2}
t_{0}   =lim_{z\rightarrow \infty}  \int_{0}^{z}\frac{dz}{H_0 (1+z)\sqrt{(\Omega_{m})_0 (z+1)^{3(\omega+1)}+(\Omega_{\Lambda})_0+(\Omega_{\sigma})_0(z+1)^6 +(\Omega_{\alpha})_{0}(1+z)^{2}}}
\end{equation}
\end{widetext}
It has been observed that the present agre of the universe is $14.53^{+0.39}_{-0.37}$ Gyrs in accordance with OHD. Similarly when we bound our model with Pantheon data then $t_{0} = 14.17^{+0.79}_{-0.72}$ Gyrs. It is important to note that the empirical values of age of the universe in WMAP observation \cite{Hinshaw/2013} is read as $t_{0} = 13.77\pm 0.059$ which is very close to the estimated value of $t_{0}$ in this paper. In some other cosmological investigations, age of the universe is estimated as $13.787\pm0.020$ Gyrs \cite{Aghamin/2018}, $14.3\pm0.6$ Gyrs \cite{Masi/2002} and $14.5\pm1.5$ Gyrs \cite{Renzini/1996}. Therefore, the universe in derived model has pretty consistency with astrophysical observations.\\  
\subsection{Jerk parameter}
The jerk parameter $(j)$, a dimensionless third order derivative of the scale factor $a(t)$ with respect
to time $t$ is an important term of cosmographic series which is used to investigate the deviations of any model of the universe from standard concordance model or $\Lambda$CDM model. It is defined as \cite{Rapetti/2007,Luongo/2013}.\\
\begin{figure}[ht!]
\begin{center}
\includegraphics[width=4cm,height=4cm,angle=0]{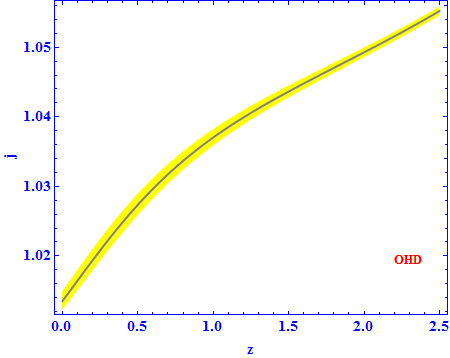}
\includegraphics[width=4cm,height=4cm,angle=0]{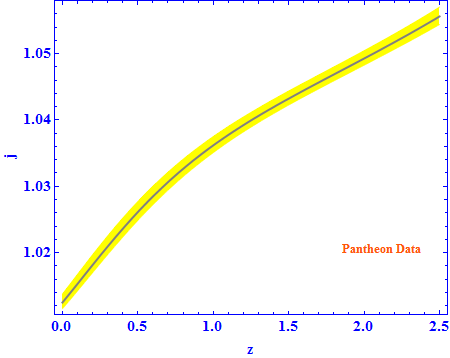}
\caption{Plot of $j$ versus $z$ for the model parameters obtained from bounding the derived model with OHD (left panel) and Pantheon data (right panel) at $68 \%$ confidence level. The solid gray line denotes the best fit curve of jerk parameter of the universe in our model.}
\end{center}
\end{figure}
\begin{equation}
\label{jerk-1}
j = \frac{\dddot{a}}{aH^{3}}
\end{equation} 
where ${\dddot{a}} = \frac{d^{3}a(t)}{dt}$.\\ 
The deceleration parameter $q$ is read as
\begin{equation}
\label{jerk-2}
q = -\frac{a\ddot{a}}{\dot{a}^{2}}
\end{equation}
Using $\dot{z} = -(1+z)H$, the reflection for $j$ in terms of $q$ and $z$ is obtained as
\[
j=q(2q+1)+(1+z)\frac{dq}{dz} = 1 - (1+z)\frac{[H(z)^{2}]^{\prime}}{H(z)^{2}}+
\]
\begin{equation}
\label{jerk-3}
\frac{1}{2}(1+z)^{2}\frac{[H(z)^{2}]^{\prime\prime}}{H(z)^{2}}
\end{equation}
Therefore, the present value of the jerk parameter is read as
\begin{equation}
\label{jerk-4}
j_{0} = q_{0}+2q_{0}^{2}+\left(\frac{dq}{dz}\right)_{z =0}
\end{equation}
The behaviour of jerk parameter $j$ versus red-shift $z$ is shown in Fig. 5. The solid gray line in the Fig. 5 represents bet fit curve of $j$ at $68\%$ confidence region. The present values of jerk parameter in our model are estimated as $j_{0} = 1.013\pm 0.001$ and $j_{0} = 1.012\pm 0.001$ in accordance with 31 OHD and Pantheon data respectively. In Akarsu et al \cite{Akarsu/2014}, the present value of the jerk parameter is estimated as $j_{0} = 1$ for the $\Lambda$CDM model. However, in this paper, $j_{0}$ is found with little deviation from its $\Lambda$CDM value which implies that our model does not behave like the standard $\Lambda$CDM universe. In the recent past, Zhai
et al \cite{Zhai/2013} have parameterized $j(z)$ phenomenologically aiming at measuring the departure of $j$ from the $\Lambda$CDM value. Later on, some other parametric reconstructions of the cosmological jerk have been investigated in different physical contexts \cite{Mukherjee/2016,Mamon/2018}. Note that the Refs. \cite{Sahni/2003,Alam/2003} deal with various DE models and role of jerk parameters to discriminate these cosmological models. Recently Singh and Nagpal \cite{Singh/2020} have investigated some values of $j_{0}$ which is not equal to $1$ by using some observational data sets.
\\
\section{Conclusion}\label{5}
In this paper, firstly, we have investigated an exact observable universe in Bianchi type V space time. Secondly, we have constrained the various model parameters of the universe in derived model by executing statistical $\chi^{2}$ tests. The main result of statistical analysis is given in Table \ref{Tab2}.\\
\begin{table}[ht]
\caption{The present values of model parameters}
\centering
\setlength{\tabcolsep}{25pt}
\scalebox{0.55}{
\begin{tabular} {cccc}
\hline
S. N. & Model parameters & $H(z)$ data & Pantheon data \\[0.4ex]			
\hline{\smallskip}
1  &  $H_{0}$ &   $67.46\pm 1.2$        & $70.02\pm1.8$   \\
			
2  &  $(\Omega_{m})_{0}$ &      $0.29\pm0.005$       & $0.273\pm0.003$    \\
			
3  &  $(\Omega_{\Lambda})_{0}$ &    $0.69\pm0.01$     & $0.727\pm0.014$  \\
			
4  &  $q_{0}$ &      $-0.59^{+0.04}_{-0.03}$        & $-0.59^{+0.02}_{-0.03}$    \\
			
5  &  $t_{0}$ &      $14.53^{+0.39}_{-0.37}$        & $14.17^{+0.79}_{-0.72}$    \\
			
6  &  $j_{0}$ &    $1.013\pm0.001$     & $1.012\pm0.001$    \\
			
\hline
\end{tabular}}
\label{Tab2}
\end{table}

The characteristics of the universe in derived model are as follows:
\begin{itemize}
\item[i)] We have obtained an exact solution of Einstein's field equations in Bianchi type V space-time rather than an adhoc parametric reconstruction of $j$ or $q$.\\

\item[ii)] The universe in derived model is able to describe the dynamics of universe in its early phase as well as the late time acceleration. We observe that there is a signature flip at transition red-shift of the derived universe. The result of this flipping, the decelerated universe turns into accelerating universe and continue to accelerate at present epoch. Also we have estimated the present value of the deceleration parameter which is in good agreement with recent astrophysical observations.\\
   
\item[iii)] The age of the universe in derived model is in good consistency with its empirical value obtained from WMAP observations and Plank collaborations.\\

\item[iv)] The analysis of the jerk parameter shows that our model has little deviation from $\Lambda$CDM universe.\\ 
\end{itemize}

As a final comment, we have investigated an exact observable universe in Bianchi type V space-time by taking into account the cosmological constant as source of energy.  The universe in derived model is a transitioning universe with transition red-shift $z_{t} = 0.74\pm0.06$ and $z_{t} = 0.72\pm0.04$ in accordance with OHD and pantheon data respectively.. It is important to note here that in Akarsu et al. \cite{Akarsu/2019}, the Bianchi I universe has been investigated as an extension of the $\Lambda$CDM model. In this paper, we have used an entirely different mechanism for solving the field equations as adopted in Ref. \cite{Akarsu/2019} and investigated an extension of the $\Lambda$CDM model in Bianchi type V space-time.\\
\section*{Acknowledgements} This work is based on the research supported in part by the National Research Foundation of South Africa (Grant Numbers: 118511).

\end{document}